\documentclass[12pt]{JHEP3}
\usepackage{graphics,psfrag}
\usepackage{amsthm,amssymb,epsfig,amsmath,euscript,array,cite}

\newcommand{\be}{\begin{equation}}
\newcommand{\ee}{\end{equation}}
\def\bea{\begin{eqnarray}}
\def\eea{\end{eqnarray}}
\newcommand{\eq}[1]{(\ref{#1})}
\def\nn{\nonumber}
\def\del{\partial}

\def\St{{\tilde{S}}}

\def\tm{{\tt m}}
\def\tn{{\tt n}}

\def\tk{{\tt k}}

\newcommand{\beq}{\begin{equation}}
\newcommand{\eeq}{\end{equation}}
\newcommand{\ben}{\begin{eqnarray}}
\newcommand{\een}{\end{eqnarray}}
\newcommand{\bes}{\begin{subequations}}
\newcommand{\ees}{\end{subequations}}
\newcommand{\blg}{\begin{align}}
\newcommand{\elg}{\end{align}}

\newcommand{\half}{{\frac{1}{2}}}
\newcommand{\quart}{{\frac{1}{4}}}

\newcommand{\Tr}{{\rm Tr}}

\newcommand{\cN}{{\cal N}}

\newcommand{\cD}{{\cal D}}

\newcommand{\cP}{{\cal P}}

\newcommand{\hgamma}{{\hat{\gamma}}}

\def\N{{\cal N}}

\def\Tr{{\rm Tr}}
\def\one{\mbox{1 \kern-.59em {\rm l}}}

\def\a{\alpha}      
\def\b{\beta}       
\def\g{\gamma}    
\def\d{\delta}    
\def\e{\epsilon}

\def\l{\lambda} \def\L{\Lambda}
\def\m{\mu} \def\n{\nu}
\def\o{\omega}

\def\s{\sigma}  
\def\t{\tau}
\def\th{\theta} \def\Th{\Theta}

\def\cD{{\cal D}}

 \def\cN{{\cal N}} 
\def\cP{{\cal P}}

\def\thb{\overline{\theta}}

\def\Phib{\overline{\Phi}}

\title{Near BPS Wilson Loop in $\b$-deformed Theories}

\author{Chong-Sun Chu, Dimitrios Giataganas
\\
Department of Mathematical Sciences, University of Durham,
Durham, DH1 3LE, UK\\
{\tt chong-sun.chu@durham.ac.uk,}\,
\tt dimitrios.giataganas@durham.ac.uk}

\abstract{
We propose a definition of the Wilson loop operator in the $\cN=1$
$\b$-deformed supersymmetric Yang-Mills theory. 
Although the operator is not BPS, it has a finite expectation value
at least up to order $(g^2 N)^2$. This does not happen generally for a  
generic non-BPS Wilson loop whose expectation value is UV divergent.
For this reason we
call this a near-BPS Wilson loop and 
conjecture that its exact expectation value is finite.
We  derive
the general form of the boundary condition satisfied
by the dual string worldsheet and find that it is deformed.
Finiteness of the expectation value of the Wilson loop,
together with some rather remarkable properties of the Lunin-Maldacena
metric and the $B$-field, fixes the
boundary condition to be one which is
characterized by the vielbein of the
deformed supergravity metric.
The Wilson loop operators provide  
natural candidates as dual descriptions to some of the 
existing 
D-brane configurations in the Lunin-Maldacena background. We also construct the
string dual configuration for a near-1/4 BPS circular Wilson loop
operator. The string lies on a deformed three-sphere instead of a
two-sphere as in the undeformed case. The expectation value of the
Wilson loop operator is computed using the AdS/CFT correspondence and is
found to be independent of the deformation. We conjecture that the exact
expectation value of the Wilson loop is given by the same matrix model
as in the undeformed case.}

\preprint{{\tt arXiv:0708.0797[hep-th]}}

\begin{document}

\setcounter{page}0

\section{Introduction}

The AdS/CFT correspondence states the equivalence of string theory on
$AdS_5 \times S^5$ to the $\cN=4$ supersymmetric Yang-Mills
\cite{adscft1,adscft2,adscft3,adscft4}.
According to this correspondence, there exists a map between gauge invariant
operators in the field theory and states in the string theory.
The correspondence is well understood for the case of half BPS local operators
where the dual string states are D-branes in the bulk
\cite{cjr,berenstein}.
The Wilson loop operator is another  important class of gauge invariant observable
which is  non-local.
The Wilson loop operator in the Euclidean $\cN=4$ SYM theory
is given by \cite{Mal}
\beq \label{wilsono}
W_R[C] =\frac{1}{N}\Tr_R\,P\,
\exp\left(\oint_C d\t (iA_\mu \dot x^\mu+\varphi_i\dot y^i)\right),
\eeq
where $A_\mu$ are the gauge fields and $\varphi_i$ are the six real
scalars.
The loop $C$ is parametrized by the variables   $(x^\mu(\t), y^i(\t))$, where
$(x^\mu(\t))$ determines the actual loop in four dimensions, and
$(y^i(\t))$ can be thought of as the extra six coordinates of
the ten-dimensional $\cN=1$ super Yang-Mills theory, of which
theory is the dimensionally reduced version.
$R$ is the representation of the gauge group $G$. In this paper we will
be interested in the case $G=U(N)$.
In \eq{wilsono}, the coupling to the gauge fields
and the scalar fields is controlled by $\dot{x}^\mu$ and $\dot{y}^i$.
In particular, Wilson loop operator satisfying the constraint
\beq\label{basicconstr}
\dot{x}^2=\dot{y}^2
\eeq
is locally BPS.
Moreover it has a finite expectation value.

In the AdS/CFT correspondence, BPS Wilson loop operators in the fundamental
representation is
dual to a fundamental string worldsheet ending on the $AdS_5$ boundary
\cite{Mal,rey}. Recently,
it has been realized that Wilson loop in the symmetric
or antisymmetric representation can be described in terms of a single
D3-brane or D5-brane with worldvolume RR flux. See
\cite{df,yamaguchi,GP,OS,HP,GP2,kumar} for the 1/2 BPS case
and \cite{dgrt} for the  D3-brane dual for 1/4 BPS Wilson in
symmetric representation. More generally, it has been shown in
\cite{GP,GP2} that half BPS Wilson loop operators in general
higher rank representations  can be described in terms of a
certain array of D3 branes or D5-branes. Analogous to the approach of
\cite{LLM},
the supergravity description
for certain half BPS Wilson loop has also been obtained
\cite{yama-bubble,lunin,gr,gutperle}.

The goal of this paper is to
try to  extend some of these results to theories with
less supersymmetries. We will consider an $\cN=1$ supersymmetric gauge theory
obtained by a marginal $\beta$-deformation of the $\cN=4$ SYM. The
theory is described by the superpotential
\beq
\label{superpot2}
i h \, \Tr( e^{ i \pi \beta } \Phi_1 \Phi_2 \Phi_3 - e^{-i \pi
  \beta } \Phi_1 \Phi_3 \Phi_2 )\ ,
\eeq
where $\Phi_i$ are the three $\N=1$ chiral superfields.
The theory is conformal provided a condition on the parameters $h,\b$ and the
gauge coupling $\tau$
is satisfied.
The resulting theory preserves $\N=1$ superconformal symmetry
and has a global $U(1)\times U(1)$ symmetry
\ben \label{charge}\nonumber
U(1)_1:& ~~~~&(\Phi_1,\Phi_2 ,\Phi_3 ) \to
(\Phi_1,e^{i\d_1}\Phi_2 ,e^{-i \d_1} \Phi_3 )
\\ \label{u1s}
U(1)_2:&~~~~&(\Phi_1,\Phi_2 ,\Phi_3 ) \to (e^{-i \d_2}
\Phi_1,e^{i\d_2}\Phi_2 , \Phi_3 ) .
\een
The $U(1)_R$ symmetry acts as
\be
U(1)_R: ~~~~(\Phi_1,\Phi_2 ,\Phi_3 ) \to
e^{i \d}(\Phi_1,\Phi_2 , \Phi_3 )
\ee
under a rotation $\th \to e^{3i \d/2}\th$.
All together, the $\cN=1$ $\b$-deformed SYM
theory is invariant under a $U(1)^3$ symmetry. It's action on the
scalar components is
\be \label{u1-sym}
\Phi_k \to e^{i \d_k} \; \Phi_k,\quad \mbox{for arbitrary
constants $\d_k$}, \quad (k=1,2,3).
\ee
Here we have used the same notation $\Phi_k$ to denote both the
lowest component of the superfield as well as the superfield itself.

The supergravity dual of the $\b$-deformed SYM was found by
Lunin and Maldacena in \cite{LM}. The Lunin-Maldacena background can
be obtained from the $AdS_5 \times S^5$ via a series of T-duality
transformation, shift and T-duality transformation acting on
the five-sphere (S-duality is also needed if $\b$ is complex).
We will look at the real $\b$ case. The supergravity description
is valid in the limit of small curvature $R = (4 \pi g_s N)^{1/4}\gg 1$ and
\be \label{largeN}
R \b \ll 1,
\ee
with
\be
R^2 \b := \hat{\g} \quad \mbox{fixed}.
\ee

Aspects of the supergravity duals of
Wilson loops
in the $\b$-deformed SYM theory  has  been studied before \cite{sfetsos,naqvi}
\footnote{The construction of \cite{naqvi}
utilizes some interesting properties found
  for giant gravitons in the Lunin-Maldacena background \cite{pm}}.
However the form of the field theory operators that are in dual with
the supergravity configurations has not been identified.
We note that  the Wilson loop operator
\eq{wilsono}, \eq{basicconstr}
is non-BPS since the gauge bosons and the scalars are in
different $\cN=1$ supersymmetry multiplets and so their supersymmetry
variations cannot cancel out each other. Conformal supersymmetry also
does not mix these multiplets.
\footnote{We note, however,
that the Wilson loop operator \eq{wilsono} is half BPS if the curve is
taken to be a lightlike line (possible in the Lorentzian case) and
with $\dot{y}^i =0 $.
This operator has no coupling to the scalar
fields and is not sensitive to the deformation.
In this paper we focus in the case where the
Wilson loop has coupling to the scalar fields since
we are interested in the effects of the $\b$-deformation.
We thanks Nadav Drukker for a discussion on this.}
One can check that even by allowing general fermion couping, it is not
possible to construct a supersymmetric Wilson loop.
It thus appears impossible to construct a Wilson loop operator
which respects some of the $\cN=1$
superconformal symmetries of the $\b$-deformed SYM.

In this paper we point out that
although the Wilson loop operator \eq{wilsono}, \eq{basicconstr} is
non-BPS
\footnote{non-BPS in the local sense. For simplicity, unless otherwise stated,
we will  omit ``local'' in the following.
The meaning should be clear from the context.
},
it shares a distinguished property of the
locally
BPS Wilson loop operator in the
$\cN=4$ theory - namely, it has a
finite vev. This is not true for a generic non-BPS
Wilson loop.
To distinguish it from a generic non
BPS loop, we call  the operator \eq{wilsono}, \eq{basicconstr} a
{\it near BPS}
{\it Wilson loop operator}.
An analogous example is the BMN operator in the
$\cN=4$ SYM theory. The BMN operator is not a BPS operator, but it has
a finite anomalous dimensions in a particular double scaling limit
\cite{BMN}. This operator is very interesting and have been studied
extensively.
We stress that the near BPS Wilson loop operator is not a deformation
of a BPS one.  The use of ``near'' is to emphasis that although it is
not  BPS, but it has finite expectation value just as a
BPS Wilson loop operator in the $\cN=4$ theory does.

We  propose that dual operators for the D-brane configurations in
\cite{naqvi} are given by
the {\it near BPS Wilson loop operators} \eq{wilsono},
\eq{basicconstr} whose path is a circle in the $x$-space and a point
in the transverse space.  When $\b \approx 0$, an
approximate half of the associated $\cN=4$ supersymmetry is
preserved. And one may call this Wilson loop operator {\it  near-half BPS}.
We also consider the  {\it near-1/4 BPS} case and
construct the dual microscopic string description. The Wilson loop's
expectation value is computed using the AdS/CFT correspondence and,
as expected, it is finite. Unlike the  near-1/2 BPS
Wilson loops where the authors find that precisely the same undeformed  ansatz
has to be taken to construct the desired dual D-branes
configurations,
here we
find that one has to employ a modified ansatz to construct the dual
string minimal surface.

The paper organized as follows.
In section 2, we review the Lunin-Maldacena background in its
original form where the deformed sphere
metric is written in the angular coordinate system.
Since the 1/4 BPS Wilson loop  necessarily involves a  non-trivial
coupling to the six real scalars field, for the purpose of using
AdS/CFT, it is more convenient   to re-express
the deformed five-sphere metric and the $B$-field in terms of
the embedding $\bf{R}^6$ coordinates. We then  point out
some very remarkable properties satisfied by the deformed metric  and the
$B$-field, which will be needed later.
%
In section 3 we
review the argument for the constraint
\eq{basicconstr} on the form of the BPS Wilson loop
and
show that similar field theory arguments lead to the same form
for the Wilson loop operator.
We also derive the general form of the
modified boundary condition for the dual string
in the Lunin-Maldacena background.
Finally, we analyze the boundary contribution arising
from the Legendre transformation of the action
and show that the finiteness of the Wilson loop vev fixes the form of
the string boundary condition.
We finish, by giving in section 4 the dual string
solution in the Lunin-Maldacena background
of a  near-1/4 BPS circular Wilson loop. Unlike the undeformed case where the
string surface is confined on a $S^2$ in the five-sphere,
the string now extends on a deformed $\St^3$. The expectation value of
the Wilson loop is computed and found to be undeformed. We conjecture
the exact expectation value of the Wilson loop is given by the same
matrix model as in the undeformed case. A number of appendices are
included. In appendix A, we derive the form of the Wilson loop in the
large $N$ limit using the phase factor associated with the infinitely massive
quark obtained from the breaking $U(N+1) \to U(N) \times U(1)$. In
appendix B, we collect some of the formula of the deformed metric
expressed in the Cartesian coordinates. The Hamiltonian-Jacobi
equation in the presence of $B$-field 
is derived in appendix C. In appendix D we show that the
1-loop corrected scalar propagator and gauge boson propagator
in the Feynman gauge remains equal. Using this result, we show that 
our near BPS Wilson loop operator is free from UV divergences up to
order $(g^2N)^2$.

\section{The Lunin-Maldacena Background}

The type IIB supergravity solution that is dual to the
$\beta$-deformation of $\cN=4$ super Yang Mills was found
in~\cite{LM}. In the string frame it is:
\bes \label{LM}
\begin{align}
	ds^2 &= R^2 \left[ds^2_{AdS_5}+  \sum_i \left(d \mu_i^2+ G
            \mu_i^2 d\phi_i^2\right)
		+ \hat{\gamma}^2 G \mu_1^2 \mu_2^2 \mu_3^2 \big(
                \sum_i d\phi_i \big)^2
		\right]\,, \label{LM-metric}\\
	e^{2\phi} &= g_s G\,,\\
	B &= R^2 \hgamma\ G\ (\mu_1^2 \mu_2^2 d\phi_1 \wedge d\phi_2
		+ \mu_2^2 \mu_3^2 d\phi_2 \wedge d\phi_3 + \mu_3^2
                \mu_1^2 d\phi_3 \wedge d\phi_1)\,,\\
	C_2 &= - 4 R^2 \hgamma\ \omega_1 \wedge (d\phi_1 + d\phi_2 + d\phi_3)\,,\\
	C_4 &= \omega_4 + 4 R^4 G\ \omega_1\wedge d\phi_1 \wedge
        d\phi_2 \wedge d\phi_3\,,
\end{align}
\ees
where  $R^4 = 4 \pi g_s N$ (in units where $\a'=1$),
\begin{equation}
	G^{-1} = 1 + \hat{\gamma}^2 (\mu_1^2 \mu_2^2 + \mu_2^2 \mu_3^2
        + \mu_3^2 \mu_1^2)\,.
\end{equation}
The parameter $\hat{\gamma}$ appearing in~\eqref{LM}
 is related to the deformation parameter $\beta$ of the gauge theory by:
\begin{equation}
	\hat{\gamma}= R^2\ \beta\,.
\end{equation}
The definition of $\o_1$ and $\o_4$ can be found in \cite{LM}.

The background has the $U(1)^3$ symmetry
\be \label{u1-phi}
\phi_k \to e^{i \d_k} \phi_k, \quad \mbox{for arbitrary constant
  $\d_k$}, \quad (k=1,2,3).
\ee
This is in correspondence with the $U(1)^3$ symmetry \eq{u1-sym} of the
$\b$-deformed SYM theory.

\subsection{Properties of the deformed metric and $B$-field}

It is convenient to introduce the
Cartesian coordinates where the deformed $\St^5$ is embedded
\ben
&&Y^1 =Y \theta^1 =Y \mu_1 \cos\phi_1,\,\, \quad
Y^4= Y \theta^4 = Y \mu_1 \sin\phi_1,\nn\\
&&Y^2 = Y\theta^2= Y \mu_2 \cos\phi_2,\,\, \quad
Y^5 = Y\theta^5= Y \mu_2 \sin\phi_2,
\label{yy}\\
&&Y^3 = Y\theta^3 = Y \mu_3 \cos\phi_3,\,\,\quad
Y^6 =Y\theta^6 = Y  \mu_3 \sin\phi_3.\nn
\een
Here $Y^2 = (Y^i)^2$ and $(\th^i)^2 =1$.
With respect to this basis, the symmetry \eq{u1-phi} is translated to
\be \label{u1-Y}
Y_1+ i Y_4 \to e^{i \d_1} (Y_1+ i Y_4),\qquad
Y_2+ i Y_5 \to e^{i \d_2} (Y_2+ i Y_5),\qquad
Y_3+ i Y_6 \to e^{i \d_3} (Y_3+ i Y_6).
\ee
The metric
\eq{LM-metric} becomes
\be
ds^2 =\frac{R^2}{Y^2}\left(\sum_{\mu=0}^3dX^\mu
  dX^\mu+ dY^2+ Y^2 d\tilde{\Omega}_5^2\right) =
\frac{R^2}{Y^2}\left(\sum_{\mu=0}^3dX^\mu  dX^\mu
+ \sum_{i=1}^6 G_{ij} dY^i dY^j \right),
\ee
where $G_{ij}$ is the embedding metric of the deformed $\St^5$.
The diagonal terms of the metric are
\ben\label{g1}
G_{ii}=\frac{1}{Y^2}(\cos^2 \phi_i + G  M_i\sin^2\phi_i), \quad
G_{i+3 \, i+3}=\frac{1}{Y^2}(\sin^2 \phi_i + G M_i
\cos^2\phi_i ), \qquad i=1,2,3,
\een
where, for convenience, we have defined the
new quantities
\ben\label{grt1}
M_1=1 + \hgamma^2 \mu_2^2 \mu_3^2,\quad
M_2 =1 + \hgamma^2 \mu_1^2 \mu_3^2,\quad
M_3=1 + \hgamma^2 \mu_1^2 \mu_2^2.
\een
The non-diagonal
elements are
\ben
&& G_{12}=\,\,\,\frac{\hgamma^2}{Y^2} G \mu_1 \mu_2 \mu_3^2 \sin\phi_1 \sin\phi_2,
\,\,\quad
G_{13}= \,\,\, \frac{\hgamma^2 }{Y^2}G \mu_1 \mu_2^2 \mu_3  \sin\phi_1 \sin\phi_3,
\, \nn
\\
&&G_{15}=-\frac{\hgamma^2}{Y^2} G \mu_1 \mu_2 \mu_3^2 \sin\phi_1 \cos\phi_2,
\quad
G_{16}=-\frac{\hgamma^2}{Y^2} G \mu_1 \mu_2^2 \mu_3 \sin\phi_1 \cos\phi_3, \,
\label{g2}\\
&&G_{23}=\,\,\,\frac{\hgamma^2}{Y^2} G \mu_1^2 \mu_2 \mu_3 \sin\phi_2\sin\phi_3,
\,\,\quad
G_{26}=-\frac{\hgamma^2}{Y^2} G \mu_1^2 \mu_2 \mu_3 \sin\phi_2 \cos\phi_3.\nn
\een
The elements $G_{45},\, G_{46},\, G_{24},\, G_{34},\, G_{56},\,
G_{35}$ differ respectively from $G_{12},\, G_{13},\, G_{15},\,
G_{16},\, G_{23},\, G_{26}$ by switching all the $\cos$ and $\sin$ in
each case.
The remaining elements are
\be \label{g3}
G_{14}=\frac{1}{2Y^2} (1-G M_1) \sin2 \phi_1,\quad
G_{25}=\frac{1}{2Y^2} (1-G M_2) \sin 2 \phi_2,\quad
G_{36}=\frac{1}{2Y^2} (1-G M_3) \sin 2 \phi_3.
\ee
In the above we have given the metric elements as a function of the
angles. For convenience, we have also recorded in the appendix B the
expressions of the metric elements as a function of $Y^i$.

Even if as expected this deformed metric is not conformally flat, it
displays some remarkable symmetries. One can check that the following
identity is satisfied
\ben
Y^i G_{ij} Y^j =1,
\een
which leads to
\ben\label{thc}
\theta^i g_{ij} \theta^j =1,
\een
where we have defined
\ben
g_{ij} := Y^2 G_{ij}.\label{ggg}
\een
The $g_{ij}$ is finite at the boundary as can be easily seen from
\eqref{g1}, \eqref{g2}, \eqref{g3}.
Another interesting property of the deformed metric is that
\ben \label{id2}
\theta^i (\partial_\a g_{ij}) \theta^j = 0,
\een
where $\del_\a$ is an arbitrary derivative.
Also we have
\be \label{id2-1}
(\del_\a \th^i) g_{ij} \th^j =0,
\ee
which  follows immediately from \eq{thc}, \eq{id2}.

The $B$-field also satisfies an interesting identity.
Writing the $B$-field as
\be
B= R^2 \hgamma\ G\  (b_1+b_2+b_3),
\ee
where
\be
b_\tm := \frac{1}{2} \e_{\tm \tn \tk}
\mu_\tn^2 \mu_\tk^2 d\phi_\tn \wedge d\phi_\tk, \quad \tm,\tn,\tk =1,2,3.
\ee
It is
\bea
&&b_{\tt 3} = Y^{-4}(Y^4 Y^5 d Y^1\wedge d Y^2 + Y^1 Y^2 d
Y^4\wedge d Y^5 + Y^1 Y^5 d Y^2\wedge d Y^4 - Y^2 Y^4 d
Y^1\wedge d Y^5), \nn\\
&&b_{\tt 2}= - Y^{-4}(Y^4 Y^6 d Y^1\wedge d Y^3 + Y^1 Y^3 d Y^4\wedge d Y^6
+ Y^1 Y^6 d Y^3\wedge d Y^4 - Y^3 Y^4 d Y^1\wedge d Y^6), \nn\\
&&b_{\tt 1}=
 Y^{-4}(Y^5 Y^6 d Y^2\wedge d Y^3 + Y^2 Y^3 d Y^5\wedge d Y^6
+ Y^2 Y^6 d Y^3\wedge d Y^5 - Y^3 Y^5 d Y^2\wedge d Y^6).
\eea
It is easy to check that the $B$-field satisfies the following identity
\be \label{id3}
B_{ik}\partial_\sigma Y^k Y^i=0.
\ee
In fact the stronger form
\ben \label{id4}
{b_\tn}_{ik}\partial_\sigma Y^k Y^i=0.
\een
holds for the individual pieces composing the
$B$-field.

In our following analysis, we will use
the properties \eq{thc}, \eq{id2-1}, \eq{id3} of the metric and the $B$-field
to study  the deformed
boundary condition for the
macroscopic string ending on the Wilson loop.
It will be interesting to see in which calculation the
results  \eq{id4} for the $B$-field will be needed.


\section{Near-BPS Wilson Loop and Twisted Boundary Condition}

\subsection{Form of the Wilson loop operator}

We start out by recalling the
arguments for the form of the Wilson loop
operator \eq{wilsono} and the constraint \eq{basicconstr}
in the original undeformed $\cN=4$ case.
Firstly,
one can examine the unbroken supersymmetry on the Wilson loop operators
\cite{dgo,kovacs,zarembo}. The Wilson loop operator is locally
supersymmetric if the constraint \eq{basicconstr} is satisfied.
A second way is from perturbation
theory. One finds that
the above constraint must be
satisfied in order for the UV-divergence to cancel out in the
expectation value of
$W$. This is easy to check in the
leading  order in $g^2N := \l$
and can be extended to arbitrary higher orders in $\l$
using arguments based on the present $SO(6)$ symmetry \cite{dgo}.
Another way to derive the Wilson loop operator is by decomposing the gauge group
$U(N+1)\rightarrow U(N)\times U(1)$ in order to use the W-bosons, that
appear from this breaking \cite{Mal,dgo}.
Finally, the constraint can also
be understood from the dual supergravity point of view \cite{dgo}. Imposing
appropriate boundary conditions and then using the Hamilton-Jacobi
equation for the minimal surface, one find that only if the constraint
\eqref{basicconstr} is satisfied can the minimal surface ends on the
boundary of $AdS_5$
and giving rises to a finite vev for the Wilson loop.
We remark that the first two methods
work for any gauge group and
any representation, while modifications will be needed in
order to generalize the third and the fourth methods to other gauge group
or higher representation.

In the $\b$-deformed theory,  as we explained in the
introduction, it appears impossible to construct a supersymmetric Wilson
loop. On the other hand, supergravity configurations have been
constructed whose dual operators would have finite vev.
We propose to study
this form of
the Wilson loop operator \eq{wilsono},
\eq{basicconstr} and that it provides the dual of the
the D-brane configurations constructed in \cite{naqvi}
\footnote{In this case, the loop is taken to be a circle in the $x$-space
and a point in the transverse space $y^i$.}.

We first give field theory arguments for the choice of this operator in the
beta-deformed theories.
First,
as in the undeformed case, one may define the Wilson loop as
the phase factor associated with the W-boson probe arising from the
breaking $U(N+1)\rightarrow U(N)\times U(1)$.  In appendix A, we
calculate the deformed $\cal{N}=~$4 Lagrangian arising from this
decomposition. The action looks quite complicated at finite $N$.
However all the $\beta$-dependence drops out in the large $N$
limit of the {\it classical} action and  the resulting
operator takes the form of \eq{wilsono}, \eq{basicconstr}.
We propose this form of the Wilson loop for any $N$.

Another field theory reason is that  if ones tries to derive
the constraint in the $\b$-deformed theory using
perturbation methods, the result at the leading order of 't Hooft
coupling $\l$ is the same as in the
undeformed theory since the propagators of the $\b$-deformed
theory are not modified. Hence the UV
pole cancels if the condition \eq{basicconstr} is satisfied, as in the undeformed
case. At higher orders of $\l$, the
$\b$-deformation breaks the $SO(6)$ invariance of the scalars and
the simple argument of the undeformed case does not hold and
anymore.
However one can check explicitly the gauge boson and scalar propagator
remains equal up to order $\l$. As a result, the UV divergence
cancels out explicitly up to order  $\l^2$  if the constraint \eq{basicconstr}
holds. The details is presented in the appendix D. We conjecture
that
the UV divergences cancel exactly
in the $\b$-deformed SYM theory.
A better understanding of perturbative properties of the beta-deformed
theory would give an answer to this problem.

This result is quite remarkable since although the
$SO(6)$ symmetry is broken by the $\b$-deformation, a $SO(6)$
invariant constraint is constructed. The same constraint is also
obtained from the SUGRA analysis performed in the next subsections
and give support to the validity of this constraint \eq{basicconstr}
and the form \eq{wilsono} of the Wilson loop operator.

We next turn to the supergravity picture for support of the form of the constraint
\eq{basicconstr} and the conjecture on the UV finiteness of the Wilson loop.
Before we do this, a
a comment is in order. In order for the Wilson loop operator to respect the
$U(1)^3$ symmetry \eq{u1-sym} of the $\b$-deformed SYM,
one need to assign a corresponding rotation
\be \label{u1-y}
y_1+ i y_4 \to e^{i \d_1} (y_1+ i y_4),\quad
y_2+ i y_5 \to e^{i \d_1} (y_2+ i y_5),\quad
y_3+ i y_6 \to e^{i \d_1} (y_3+ i y_6),
\ee
to the loop variables $y_i$. Here
we have used the identification of the scalar fields \eq{sdef}.
The transformation properties \eq{u1-y} and \eq{u1-Y} leads one to
associate $y_i$ with $Y_i$. This fact is important as, given a
specific configuration of the loop variables $y_i$ in the field theory, it
tells which $Y_i$ should be activated for the dual
string configuration in supergravity. An example will be shown  in section 4.

\subsection{Deformed boundary conditions}

Since  the constraint \eq{basicconstr} is closely
related with the boundary conditions of the dual string we will use it to
analyze how these boundary conditions are modified and we will see
that they are modified for the directions in the $\St^5$.
We will first derive the most general form of the boundary
condition for the string minimal surface. This is given in terms of an
arbitrary matrix $\L^k{}_m$.
Then we  show that the field theory constraint is obtained if this matrix
is given by the vielbein of the deformed metric.
We also show that the UV divergence in the supergravity result
is cancelled.

Let $(\s_1,\s_2) =(\t,\s)$ be the worldsheet coordinates 
\footnote{Note that the conjugate momentum is defined with $\s_2 = \s$
  taken as the Euclidean time. We have chosen to denote the boundary
  coordinate $\s_1$ by $\tau$ so as to conform to the notation 
  in \eq{wilsono} which is commonly adopted.
}.
The complex structure ($\a,\b =1,2$) on the worldsheet
\be
J_\a{}^\b = \frac{1}{\sqrt{g}} g_{\a\g}\e^{\g\b}
\ee
is given in terms of the induced metric $g_{\a\b}$.
For the Lunin-Maldacena background, the Hamilton-Jacobi equation  takes the form
\ben\label{HJ0}
G^{ij}(P_i -  i B_{ik} \partial_1 Y^k) (P_j -  i B_{jl} \partial_1
Y^l) + G^{\m\n} P_\mu P_\nu
 = G_{ij} \partial_1 Y^i \partial_1  Y^j+
 G_{\mu\nu}\partial_1 X^\mu \partial_1  X^\nu
\een
where the momentum are
\be
P_{i}= G_{ij} J_1{}^\b \partial_{\b}Y^{j} + i B_{ik} \partial_1 Y^k, \quad
P_\mu =  G_{\m\n}J_1{}^\b \partial_\b X^\nu. \label{moment}
\ee 
The derivation of the HJ equation is given in the appendix. 
Notice the difference between the undeformed case is that now appears
the antisymmetric field $B_{ij}$, which is not zero in the deformed
Lagrangian. Furthermore, because we use Euclidean world-sheet,
it appears as usual an $i$ in front of the worldsheet coupling to the
$B$-field. However,
the terms including the antisymmetric field will disappear when we
substitute in the Hamilton-Jacobi equation the conjugate momentum and
we obtain
\ben
g_{ij}  J_1{}^\a J_1{}^\b \partial_{\a}Y^{i}  \partial_{\b}Y^{j}
+  J_1{}^\a J_1{}^\b \partial_{\a}X^{\m}  \partial_{\b}X^{\m}
 = g_{ij} \partial_1 Y^i \partial_1  Y^j +
( \partial_1  X^\m)^2,  \label{hj}
\een
where we have substituted \eq{ggg} and using that $G_{\m\n} =\d_{\m\n}/Y^2$.

Now let us determine the boundary conditions for the string
coordinates. Suppose that the Wilson loop is parametrized by the
values $(x^\mu(\sigma_1),y^i(\sigma_1))$ and choose the world-sheet coordinates such
that the boundary is located at $\sigma_2=0$.
Since the deformation in the dual supergravity background does not appear in
the  $X^\m$ directions,
it is natural to impose the same  Dirichlet boundary condition for
these coordinates
as in the undeformed case:
\be
X^\mu(\sigma_1,0)=x^\mu(\sigma_1).
\label{dirichlet}
\ee

For the remaining 6 string coordinates $Y^i(\sigma_1,\sigma_2)$, one can
expect the situation to be more complicated since in the Lunin-Maldacena background,
the deformations from the standard $AdS$ background occur in these directions.
Due to the presence of the $B$-field,
the general mixed  boundary condition takes the form
\ben
J_1^\a \partial_\a Y^k(\sigma_1,0) 
+ i B^{k}{}_{l} \partial_1 Y^l(\sigma_1,0) =
\Lambda^k{}_{l}\, \dot{y}^l(\sigma_1)  \label{bci}
\een
for some invertible matrix $\Lambda^k{}_{l}$. In addition, for  a
minimal surface to terminate at the boundary of $AdS_5$, we have
the  Dirichlet conditions
$Y^i(\sigma_1,0)=0$, which means
\be \label{dzero}
\partial_1 Y^i(\sigma_1,0) =0.
\ee
So the above Neumann boundary condition simplifies to
\ben
J_1^\a \partial_\a Y^k(\sigma_1,0)= \Lambda^k{}_{l}\,
\dot{y}^l(\sigma_1).
\label{bc}
\een
Inserting the boundary conditions \eq{dirichlet}, \eq{dzero} and \eq{bc}
in the Hamilton-Jacobi equation we find
 \ben
\dot{x}^2
-\Lambda^{k}{}_{m}\Lambda^{l}{}_{n} g_{kl}\, \dot{y}^m \dot{y}^n=
(J_1{}^\a \partial_\a X^\mu)^2 .
\een
The term $(J_1{}^\a \partial_\a X^\mu)^2$ has to be zero near a smooth
boundary, otherwise it costs infinite area.
Therefore, we arrived at the constraint
\be \label{gen-cond}
\dot{x}^2 =  g_{kl} \Lambda^{k}{}_{m}\Lambda^{l}{}_{n}\, \dot{y}^m \dot{y}^n.
\ee
In particular, the
constraint derived from
supergravity agrees with the constraint \eq{basicconstr} derived from field
theory considerations of the condition if the matrix $\L^k{}_{i}$
satisfies the condition
\beq\label{condition}
 g_{kl} \Lambda^k{}_{m} \Lambda^{l}{}_{n} = \delta_{mn} .
\eeq
This means that the {\it boundary condition matrix}
$\Lambda^{k}{}_{m}$ is the vielbein of the
deformed metric $g_{kl}$. We remark that in \cite{naqvi}, the D-brane boundary
condition in the $\b$-deformed theory was obtained out using TsT
transformation on the original undeformed boundary condition. It was
easy in that case since only angles was involved. In our case we
still  expect that one can perform a TsT-transformation on the
angles to derive the modified boundary condition \eq{bc},
\eq{condition}, although it is less direct since
the boundary  condition is formulated in terms of the
Cartesian coordinates while TsT transformations operates on the
angles.


\subsection{Legendre transformation and boundary contribution}

In the undeformed case, after performing a Legendre transformation the
UV singularity of the area functional cancels because of the constraint
$\dot{x}^2=\dot{y}^2$. For the deformed case, we should have a similar
situation in order for our result to be consistent. We will check this
now. As before, since the boundary condition \eq{bc} is of Neumann type,
we consider the same Legendre transform
\beq
\tilde A =A-\oint d\sigma_1 P_i Y^i.
\label{naction}
\eeq
Since the metric is singular at $Y=0$, we
introduce a regulator $Y=\e$ and  evaluate the
regularized action for $Y\geq\epsilon$.
Let us first focus on the term that comes from Legendre transformation
Using the definition \eq{moment}, we have, at $Y=\e$,
\be
P_i Y^i = G_{ij} J_1{}^\b \partial_\b Y^j Y^i = \frac{1}{Y} J_1^\a \del_\a Y ,
\ee
where we have used the property \eq{id3} to get rid of the $B$-field
term in the first equality; and substituted $Y^i =Y \theta^i $,
$G_{ij}=g_{ij}/Y^2$  and used \eq{thc},
\eq{id2-1} in the second equality. To express $J_1^\a \del_\a Y$ in terms
of the boundary data, we note on substituting  $Y^i =Y \theta^i $ and using again
\eq{thc},
\eq{id2-1} that,
\be
g_{ij} (J_1^\a \partial_\a Y^i)  (J_1^\b \partial_\b Y^j)  =
(J_1^\a \partial_a Y)^2
+ Y^2 J_1^\a J_1^\b g_{ij} \del_\a
\th^i \del_\b \th^j.
\ee
In the limit $\e \to 0$, the second term on the RHS is zero.
As for the LHS, we  use the boundary condition \eq{bc} and the condition
\eq{condition}. Therefore
we obtain
\ben
\dot{y}^2=(J_1^\a \partial_a Y)^2 .
\een
And the action \eqref{naction} becomes
\beq
\tilde A
=A-\oint d\sigma_1 \frac{|\dot{y}|}{Y} =A - \frac{1}{\epsilon}\oint
d\sigma_1 |\dot{y}| ,
\label{nfaction}
\eeq
where we are evaluating the
regularized action for $Y\geq\epsilon$.

Now, as in the undeformed case, we expect the area of the minimal surface
to have a linear divergence proportional to
the circumference of the boundary. Therefore
\beq
\tilde A
= \frac{1}{\epsilon}\oint d\sigma_1 (|\dot{x}|-|\dot{y}|)+ {\rm finite\,
part}.
\label{nnnaction}
\eeq
This  means that  like the undeformed case, the
linear divergence in the deformed case cancels when
the conditions \eqref{basicconstr} is satisfied.

It is worth noticing that this analysis of the absence of UV divergence
in the vev of the Wilson loop is valid for large $\l$, while the field
theory analysis presented in the last subsection is valid for small
(up to second order in)
$\l$. The fact that the UV divergence cancels and a
well-defined Wilson loop is obtained for both small and large $\lambda$ leads
us to the conjecture that the Wilson loop \eq{wilsono}, \eq{basicconstr}
is well-defined and has finite vev in the $\cN=1$ $\b$-deformed SYM
theory.

\section{Near-1/4 BPS Wilson Loop}

In the above, we
have proposed that the D-brane
configurations considered in \cite{naqvi} are dual to the near-1/2 BPS
operators where the circular loop has a trivial dependence in the
transverse space. Now we look at next non-trivial case where the loop
involves a non-trivial rotation in the transverse space as well,
\be \label{w-circular}
W[C] =\frac{1}{N}\Tr\,P\,
\exp\left[\int d\t
\Big(iA_\mu \dot x^\mu(\t) + |\dot{x}(\t)| \varphi_i \th^i(\t) \Big)\right],
\ee
where the loop is  a circular path of radius $R_0$ in space
\be
x^1 = R_0 \cos \t, \quad x^2 =R_0 \sin \t, \label{path1}
\ee
and the coupling to the three scalars $\varphi_1,\varphi_2,\varphi_5$ is
parametrized by
\be \label{path2}
\th^1 =\cos \th_0,\quad
\th^2 = \sin \th_0 \cos \t, \quad \th^5 = \sin \th_0 \sin \t,
\ee
with an arbitrary fixed $\th_0$. This operator in the undeformed
theory is 1/2 BPS when $\th_0 = 0$  and 1/4 BPS in general \cite{df}.
In this section we use the AdS/CFT correspondence to compute the value for the
circular near BPS Wilson loop operator in the $\b$-deformed SYM.

We use the following form for the (Euclidean) $AdS_5$ metric
\beq\label{m14}
ds^2 = du^2 + \cosh^2 u (d\rho^2 + \sinh^2\rho d\psi^2) + \sinh^2u(
d\chi^2 + \sin^2\chi d\phi^2).
\eeq
For the deformed  $\St^5$ \eq{LM-metric}, we parametrize
the $\mu_i$ coordinates via
\ben \label{mu}
\mu_1=\cos \theta,\quad \mu_2=\sin\theta \cos \alpha, \quad \mu_3=
\sin\theta \sin \alpha
\een
so that $\sum d \mu_i^2 = d\th^2 + \sin^2 \th d \a^2$. For Euclidean space,
the worldsheet coupling to the $B$-field get an extra factor of $-i$.

To find the dual string configuration, we note that
\bea
\th^1+ i \th^4  &=& \cos{\th_0} ,\nn\\
\th^2+ i \th^5 &=& \sin{\th_0} e^{i \tau},\\
\th^3+ i \th^6 &=& 0. \nn
\eea
Comparing with the definition \eq{yy} for $\th^i$, and using \eq{mu},
this means the dual string configuration must satisfy
$\phi_2= \tau, $ and $\th=\th_0$, $\phi_1 = \a =\phi_3 =0$ at the boundary.
Minimally, one wants to consider an ansatz involving
only two angles $\phi_2$ and $\th$. However due to the $B$-field, one
can see easily that this is inconsistent.
Let us therefore consider a motion
on $R^2 \times \St^3$ where
$R^2 \subset AdS_5$ is parametrized by 
$\psi$ and $\rho$, and the deformed 3-sphere is
parametrized by the three angles $\th, \phi_1, \phi_2$ with $\a=\phi_3 =0$. The
Polyakov action for the Euclidean worldsheet  $(\s,\t)$ is
\bea
S&=&\frac{\sqrt{\lambda}}{4 \pi }\int d \sigma d\tau
\bigg[  \rho'{}^2 + \dot \rho^2+
\sinh^2 \rho (\psi'{}^2+ \dot\psi^2 )+ \th'{}^2 +\dot\th^2
+G \cos^2\theta(\phi_1'{}^2+\dot\phi_1^2)  \nn\\
&& \qquad\qquad \qquad \quad + G \sin^2\theta  (\phi_2'{}^2+\dot\phi_2^2)
-2i \hat{\gamma}G \sin^2 \theta \cos^2 \theta
(\dot\phi_1 \phi_2{}' - \phi_1{}' \dot\phi_2   ) \bigg] , \label{s-action}
\eea
where $'$ (resp. $\dot{}\;$) denotes $\partial_\s$ (resp. $\partial_\t$)
derivative. 
Due to the extra factor of $-i$ in the $B$-field coupling, 
a real configuration is possible only if one perform a Wick rotation
$\phi_1 \to i \phi_1$. To match with the path
specified by \eq{path1}, \eq{path2}, we look for solution of the form
\bea
&& u=0,\quad \rho=\rho(\sigma),\quad \psi=\tau \label{a1} \\
&& \theta=\theta(\sigma), \quad \phi_1=\phi_1(\sigma),\quad \phi_2=\tau. \label{a2}
\eea
We remark that, compared to the solution
\cite{d-f1} for the undeformed case, our ansatz has an additional angle
$\phi_1$ turned on. This is similar to the situation in the story of magnon. There
the string configuration dual to the magnon was
found \cite{cgv} to expand from a motion on $S^2$ for
the undeformed case to a motion on a deformed 3-sphere when the
$\b$-deformation is turned on. We also remark that the Wick rotation
on $\phi_1$ is natural and
is consistent with a
semi-classical interpretation of the AdS/CFT correspondence as a
tunnelling phenomena.

The classical equations of motion for our ansatz \eq{a1}, \eq{a2}
takes the form
\bea
 \rho'' &=& \cosh \rho \sinh \rho, \label{1}\\
\theta'' &= &\half \partial_\theta
(G \sin^2 \theta)
-\half \partial_\theta(G \cos^2\theta)\; \phi_1'{}^2
+ \partial_\theta(\hat{\gamma} G \sin^2 \theta\; \cos^2 \theta)
\phi_1'\; , \label{2}\\
0 &=&\partial_\tau (- G \cos^2\theta\; \dot\phi_1
- \hat{\gamma} G \sin^2 \theta\; \cos^2 \theta\; \phi_2{}')+
\partial_\s( -G \cos^2 \theta\; \phi_1{}'
+ \hat{\gamma} G \sin^2 \theta\; \cos^2 \theta\; \dot\phi_2{}
)   \label{3}\\
0 &=&\partial_\tau (G \sin^2 \theta\; \dot\phi_2
+ \hat{\gamma} G \sin^2 \theta \; \cos^2 \theta\; \phi_1{}')+
\partial_\sigma (- G \sin^2\theta \;\phi_2{}'
-  \hat{\gamma} G \sin^2 \theta\; \cos^2 \theta\; \dot\phi_1)\ \label{4}.
\eea
The equation \eqref{4} is satisfied trivially. Equation \eqref{3} gives
\ben
-G \cos^2 \th \phi_1{}' + \hat{\gamma} G \sin^2 \theta
\cos^2 \theta= c_1 .
\een
For the surface to be closed, it must be possible to reach $\th=0$ (north pole)
or  $\pi$ (south pole), and there the derivatives $\phi_1{}',\phi_2{}'$ should be
zero since no rotation is possible. Therefore $c_1=0$ and we have
\ben\label{theta2}
\phi_1{}'= \hat{\gamma} \sin^2\theta .
\een
Equation \eqref{2} then becomes
\ben\label{thetafin}
\theta'^2=\sin^2 \theta + c_2,
\een
where $c_2$ is a constant. Notice how the $G$ dependence
disappears in the above calculations. Finally, we check also the
Virasoro constraints, which reads
\be\nonumber
 \rho'{}^2- \sinh^2 \rho + \theta'{}^2 - G
\sin^2 \theta -G \cos^2 \theta \; \phi_1'{}^2=0 ,
\ee
which implies
\be\label{virasoro2}
- \rho'{}^2+ \sinh^2 \rho = \theta'{}^2 - \sin^2 \theta .
\ee
Again here notice that the $G$ dependence
disappears.  To get a surface in correspondence to a single circle,
we set $c_2 =0$,
and the final form of the equations of motion is
\ben\label{eomf}
\rho'^2&=& \sinh^2 \rho, \\
\theta'^2&=& \sin^2 \theta .
\een
This give the solution
\ben\label{rho}
\sinh \rho= \frac{1}{\sinh \sigma},
\een
\ben\label{th}
\sin\theta=\frac{1}{\cosh(\sigma_0 \pm \sigma)}\, \Leftrightarrow \,
\cos \theta = \tanh(\sigma_0\pm \sigma)
\een
and
\be
\phi_1 = \hat{\gamma} \big(\tanh(\s \pm \s_0) \mp \tanh(\s_0) \big).
\ee

To see how our solution behaves, consider the limits
\bea
\sigma\rightarrow 0 &\Rightarrow& \rho\rightarrow \infty,
\quad\mbox{and} \quad
\th\rightarrow \th_0,\quad
\phi_1 \to 0,
\\
\sigma\rightarrow \infty &\Rightarrow& \rho\rightarrow 0,
\;\;\quad\mbox{and}\quad
\theta\rightarrow \mbox{0 or $ \pi$}.
\een
Here $\cos \theta_0 =  \tanh\sigma_0 $.
Depending on the sign in \eq{th}, the surface extends over the north or south pole
of $\St^5$.

Next we evaluate the action for this configuration. The bulk term is
\ben
S_{\rm bulk}= \frac{\sqrt{\lambda}}{2 \pi }\int d \sigma d\tau  ( \sinh^2 \rho +
\sin^2\theta),
\een
from which we find 
\ben
S_{\rm bulk} =\sqrt{\lambda} ( \coth \rho_{\rm max} \mp \cos\theta_0).
\een
Here we have introduced a cutoff $\s_{\rm min}$ to regulate the boundary
contribution, and 
$\rho_{\rm max}$ is the corresponding cutoff on $\rho$.
The $\coth \rho_{\rm max}$ term
will cancel with boundary term  coming from the Legendre transformation
as we have showed above.
Hence, the final result is
\ben
S_{tot}=\mp \sqrt{\lambda} \cos\theta_0,
\een
and
\be
\langle W \rangle \sim \exp \big(\pm \sqrt{\lambda} \cos\theta_0\big),
\ee
where the sign is we chosen to minimize the action.
This is the same vev as the 1/4 BPS Wilson loop in the
undeformed theory.

We note that in addition to this supergravity solution which involves
3 angles, one can also construct a solution which involves only the
two angles
\be
\th= \th(\s), \quad \a =\tau,
\ee
together with \eq{a1}.
This solution is exactly the same as the undeformed one given in
\cite{d-f1} and gives rises to the same expectation value for the dual
Wilson loop.
It is straightforward to work out the Wilson loop operator that is
dual to it. It is defined by the loop
\be \label{path2'}
\th_1 = \cos \th_0, \quad \th_2 = \sin \th_0 \cos \tau, \quad
\th_3 = \sin \th_0 \sin \tau.
\ee
Due to a lack of $SO(6)$ invariance, the Wilson  loop operator with the
loop \eq{path2'} is different from the one with the loop
\eq{path2}. It is quite amazing that they have the same expectation
value.

To understand this result better.
Let us first recall how the expectation value of the 1/2 BPS circular Wilson
loop was computed in gauge theory \cite{ESZ,dg}. The
circular loop is related to the straight line by a conformal transformation,
one can therefore relate the circular Wilson loop to the expectation
value of the Wilson straight
line, which is one. The result is however non-trivial since under the conformal
transformation, the gluon propagator is modified by a singular total
derivative which gives non-zero contribution only when both ends of the
propagator are located at the point which is conformally
mapped to the infinity. It was conjectured by \cite{ESZ} that diagrams
with internal vertexes cancel precisely and this is supported by a
direct calculation at order $g^4 N^2$.
Assuming this is true, \cite{dg} showed that the sum
of all the non-interacting diagrams can be written as
a Hermitian matrix model
\be
\langle W_R \rangle = \Big\langle \frac{1}{N}\Tr_R\big[ e^M\big] \Big\rangle =
\frac{1}{Z} \int \cD M \frac{1}{N}
\Tr_R\big[ e^M\big] \exp\big(-\frac{2N}{\l} \Tr M^2\big).
\ee
This is exact to  all order in $\l$ and $1/N$ \cite{dg}. Explicit
evaluation of the integral and hence the Wilson loop expectation
value has been performed for loops in
various representations \cite{ESZ,dg,df,yamaguchi,OS,HP}.
This argument has also been applied to the 1/4 BPS fundamental Wilson
loop \cite{d-f1}.

Now the $\b$-deformed theory is exact conformal. So the above argument
of conformal anomaly applies. The only thing one need to be sure is how
interacting diagrams contribute. If they again sum up to zero, then
there is no $\b$-dependence left and one will get the same result as in
the undeformed case. Our result of getting the same expectation value
for the undeformed and the deformed Wilson loop operators
suggests that the interacting
diagrams again cancel
exactly, at least in the large 't Hooft coupling limit.
This is however not easy to
prove from perturbation theory since one needs to identify terms with
dependence on $\b^2 N$ at each order of $1/N$. We believe
a similar
mechanism as in the undeformed case is at work. If this is the case, the
exact expectation value of the circular Wilson loop in the $\b$-deformed
SYM will be given by the same matrix model as in the undeformed $\cN=4$
case. A better understanding of how this works in the undeformed case is
necessary and will be very interesting.

For the same reason, we conjecture that the expectation value of the
near-1/4 BPS Wilson loop in higher representations will also be
unmodified. It will be interesting to construct the D3-brane and
D5-brane dual to these Wilson loops in higher representations for the
$\b$-deformed theory and check this.

In this paper we have proposed a definition of a near BPS Wilson loop
operator in the $\b$-deformed SYM theory.
We conjectured that this operator has finite vev and
provided supporting evidences both from field theory and from supergravity.
Thus this operator is a natural
candidate of a Wilson loop operator which admits a holographic
description in the $\beta$-deformed AdS/CFT correspondence. We showed
that on the supergravity side, the finiteness of the vev of the Wilson
loop implies the same constraint on the loop as is derived from the
field theory analysis. That this is true relies on some remarkable
properties satisfied by the metric and the $B$-field of the
Lunin-Maldacena background. It will be interesting to be able to
formulate and understand these symmetry properties in terms of the dual
field theory language. Its origin is likely to be nonperturbative. This
should provide us a better understanding of the mechanism responsible
for the
finiteness of the vev of the Wilson loop.


\section*{Acknowledgements}

We thank
Nadav Drukker for  useful email exchange and discussions.
The research of CSC is  supported by EPSRC and PPARC.
The work of DG is supported by an EPSRC studentship.

\newpage
\appendix

\section{Wilson loop from  $U(N+1)\rightarrow U(N)\times
U(1)$ breaking}

For real $\b$-deformation, the bosonic part of the
Lagrangian of the $\beta$-deformed SYM theory is given by
\ben
&&{\cal{L}} = \, \Tr \Bigg( {\frac{1}{4}}
F^{\mu \nu}F_{\mu \nu} +
(D^\mu \bar \Phi^\a ) (D_\mu \Phi_\a  )
- {g^2} [\Phi_\a,\Phi_\b]_* [\bar \Phi^\a,\bar \Phi^\b]_*
+{\frac{g^2}{2}}[\Phi_\a,\bar \Phi^\a][\Phi_\b,\bar \Phi^\b] \Bigg),
\label{Ldef}\quad
\een
where $\Phi_\a$ ($\a=1,2,3$) are the scalar components of the $\cN=1$
chiral superfield.
The star product for the fields is defined by
\be
\label{star}
f * g :=
e^{ i  \pi \beta (Q_1^f Q_2^g - Q_2^f Q_1^g) } f g ,
\ee
where $fg$ is an
ordinary product
and $(Q_1^{\rm field},Q_2^{\rm field})$ are the $U(1)_1 \times U(1)_2$
charges of the fields ($f$ or $g$).
The values of the charges for all fields are given in \eqref{charge}.
Clearly the star product is non-trivial only when different chiral fields are
multiplied, as explicit in \eq{Ldef}.
Furthermore, we use the deformed commutator of fields,
\ben
[ f_\alpha,g_\gamma]_{*} :=\, f_\alpha * g_\gamma - g_\gamma * f_\alpha = \,
e^{ i \pi \beta_{\alpha \gamma} }\, f_\alpha g_\gamma-\,
e^{ -i \pi \beta_{\alpha \gamma} }\, g_\gamma f_\alpha \ , \label{combeta}
\een
where $\beta_{\alpha \gamma}$ takes the values
\ben
\beta_{\alpha \gamma}=-\beta_{\gamma \alpha } \, , \quad
\beta_{12}=\, -\beta_{13}=\, \beta_{23} :=\, \beta \ .
\label{betaijs}
\een

Next, let us break the gauge group $U(N+1)\rightarrow U(N)\times
U(1)$ by turning non-zero vacuum expectation values for the scalar
fields
\be
\Phi_\a = \left(\begin{array}{cc} 0_{N\times N} & 0 \\ 0 &  M \Th_\a \end{array}
\right),\quad \a=1,2,3 \ .
\ee
Here  $\Th_\a$ lies on a 5-sphere,
$\Th_\a \Th^\a=1$, corresponding to the direction of the symmetry breaking.
Decomposing the fields as
\beq \hat A_\mu = \left(\begin{array}{cc}
A_\mu & W_\mu \\
 W_\mu^\dagger&a_\mu\end{array}\right),
\qquad \hat \Phi_\alpha = \left(\begin{array}{cc}\Phi_\alpha & W_\alpha \\
Y_\alpha&M\Theta_\alpha\end{array}\right),
\eeq
we obtain the action in terms of $W_\alpha,\,Y_\alpha$ :
\ben\label{ulagr}
\nonumber
&&
\hat S=
\quart F_{\mu\nu}^2
+ (D_\mu \overline{\Phi}_\alpha)(D_\mu \Phi_\alpha)
+ \half[\Phi_\alpha,\,\overline{\Phi}_\alpha][\Phi_\gamma,\,\overline{\Phi}_\gamma]
+ [\Phi_\alpha,\,\Phi_\gamma]_{\beta_{\alpha
    \gamma}}[\overline{\Phi}_\alpha,\,\overline{\Phi}_\gamma]_{\beta_{\alpha \gamma}}
\nonumber\\
&&
+\big((D_\mu-ia_\mu) W_\alpha^\dag\big)\big((D_\mu+i a_\mu) Y_\alpha\big)+
\big((D_\mu+i a_\mu) Y_\alpha^\dag\big)\big((D_\mu-i a_\mu) W{_\alpha}\big)
\nn\\
&& + \quart f_{\mu\nu}^2 +(\partial_\mu
M\Theta_\alpha^\dag)(\partial_\mu M\Theta_\alpha)
\nonumber\\
&&
-2Y_\alpha^\dagger\Big(\Phi_\gamma\Phi_\alpha e^{-2 i \pi \beta_{\alpha
    \gamma}}-\Phi_\alpha\Phi_\gamma
+\half(\Phi_\alpha-M\Theta_\alpha)(\Phi_\gamma-M\Theta_\gamma)+ M^2
\Theta_\alpha\Theta_\gamma(e^{2  i \pi \beta_{\alpha
    \gamma}}-1)\Big)W_\gamma^\dag
\nonumber\\
&&
-2Y_\alpha\Big(\overline{\Phi}_\gamma\overline{\Phi}_\alpha e^{-2 i \pi
  \beta_{\alpha \gamma}}-\overline{\Phi}_\alpha\overline{\Phi}_\gamma
+
\half(\overline{\Phi}_\alpha-M\overline{\Theta_\alpha})
(\overline{\Phi}_\gamma-M\overline{\Theta}_\gamma)+
M^2 \overline{\Theta}_\alpha\overline{\Theta}_\gamma(e^{2 i \pi
  \beta_{\alpha \gamma}}-1)\Big)W_\gamma
\nonumber\\
&&
+Y_\alpha^\dagger\Big(
\big(2(\overline{\Phi}_\kappa-M\overline{\Theta}_\kappa)**_{\beta_{\alpha
    k}}(\Phi_\kappa-M\Theta_\kappa)+
[\Phi_k,\overline{\Phi}_\kappa]\big)\delta_{\alpha \gamma}
-2(\overline{\Phi}_\gamma-M
\overline{\Theta}_\gamma)*_{\beta_{\alpha \gamma}}(\Phi_\alpha-M\Theta_\alpha)
\nonumber\\
&&
\qquad\qquad\qquad\qquad\qquad\qquad\qquad\qquad+
(\Phi_\alpha-M\Theta_\alpha)(\overline{\Phi}_\gamma-
M\overline{\Theta}_\gamma)\Big)
W_\gamma
\nonumber\\
&&
+Y_\alpha\Big(\big(
2(\Phi_\kappa-M\Theta_k)**_{\beta_{\alpha k}}(\overline{\Phi}_\kappa-M
\overline{\Theta}_\kappa)+[\overline{\Phi}_\kappa,\Phi_\kappa]\big)\delta_{\alpha
  \gamma}
-2(\Phi_\gamma-M\Theta_\gamma)*_{\beta_{\alpha \gamma}}
(\overline{\Phi}_\alpha-M\overline{\Theta}_\alpha)
\nonumber\\
&&
\qquad\qquad\qquad\qquad\qquad\qquad\qquad\qquad+
(\overline{\Phi}_\alpha-
M\overline{\Theta}_\alpha)(\Phi_\gamma-M\Theta_\gamma)
\Big)W_\gamma^\dag+ \cdots \ ,
\een
where we have defined
\bea
\nonumber
&&(\overline{\Phi}_\kappa-M
\overline{\Theta}_\kappa)**_{\beta_{\alpha \kappa}}
(\Phi_\kappa-M\Theta_\kappa):=
\overline{\Phi}_\kappa \Phi_\kappa
+ M^2 \overline{\Theta}_\kappa \Theta_\kappa
-\overline{\Phi}_\kappa M \Theta_\kappa e^{2 i\pi \beta_{\alpha \kappa}}
- M\overline{\Theta}_\kappa \Phi_\kappa e^{- 2 i\pi \beta_{\alpha
\kappa}} \ ,
\nonumber\\
&&
(\overline{\Phi}_\gamma-M \overline{\Theta}_\gamma)*_{\beta_{ \alpha \gamma}}
(\Phi_\alpha-M\Theta_\alpha):=
\overline{\Phi}_\gamma \Phi_\alpha e^{2 i\pi \beta_{\alpha \gamma}}
+ M^2\overline{\Theta}_\gamma \Theta_\alpha e^{-2 i\pi \beta_{\alpha \gamma}}
-M\overline{\Theta}_\gamma \Phi_\alpha
- \overline{\Phi}_\gamma M\Theta_\alpha. \nn
\eea
In \eq{ulagr}, $\cdots$ denotes terms of higher order (fourth) in the fields $W, Y$,
and  $\Tr$ over $U(N)$ is understood.

Next, we go to the real basis by introducing
\bea \label{sdef}
\Phi_1 = \, \frac{1}{\sqrt{2}} (\varphi_1 +i\varphi_4),\quad
\Phi_2 &=& \, \frac{1}{\sqrt{2}} (\varphi_2 +i\varphi_5),\quad
\Phi_3 = \, \frac{1}{\sqrt{2}} (\varphi_3 +i\varphi_6)\ ,
\label{phisrel} \\
 W_1 = \, \frac{1}{\sqrt{2}} (w_1 +i w_4),\quad
W_2 &=& \, \frac{1}{\sqrt{2}} (w_2 +i w_5),\quad W_3 = \,
\frac{1}{\sqrt{2}}
(w_3 +i w_6)\ ,
\label{dictphi}
\eea
and similarly for $Y_\a$ and $\Th_\a$.
The terms $Y^\dag_\alpha(\cdots)W^\dag_\gamma$,
$Y_\alpha(\cdots)W_\gamma $, $Y_\alpha(\cdots)W^\dag_\gamma$
and $Y^\dag_\alpha(\cdots)W_\gamma$
become
\bea\label{wfinal}
\nonumber
&&
\sum_{i=1}^6 w_i^\dag
\Big[ \sum_{j=1}^6 C_{jj}- C^0_{ii} \Big]w_i \nn\\
&&+ \sum_{ij=14,25,36}
 w_i^\dag \Big[2\lambda_{ij} - C^0_{ij}
+2 i \sin 2\pi\beta \sum_{kl}s_{ikl}\varphi_k M \theta_l\Big]w_j + ~c.c.\\
&& +\sum_{\begin{subarray}{l}
ij\neq 14,25,36\\
i\neq j
\end{subarray}}
w_i^\dag\Big[2 \Lambda_{ij} - C^0_{ij}
- 2 M^2 \theta_i \theta_j (\cos 2\pi\beta -1)
+ 2 i \sum_{k,l=1}^6 S_{iklj} \sin 2\pi\beta (\varphi_k \varphi_l
- M^2 \theta_k \theta_l)\Big]w_j \nn\\
&&\qquad\qquad\quad + ~c.c., \nn
\eea
where we have defined
\bea
&&C_{ij}:=
(\varphi_i e^{i \pi \b} -M \theta_i e^{-i \pi \b})(\varphi_j e^{-i \pi \b}
-M\theta_j e^{i \pi \b}) , \\
&&\Lambda_{ij}:=\varphi_i \varphi_j-\varphi_j\varphi_i \cos 2 \pi \beta \ ,\\
&&\lambda_{ij}:=[\varphi_i, \varphi_j]\ ,
\eea
and $C^0_{ij} = C_{ij} (\b=0)$.
The quantities $s_{ijk},\,S_{ijkm}$ are equal to $\pm 1$ or zero,
and their non-zero elements are shown below:
\ben
&&s_{ikl}=1 \quad \mbox{for $ikl= 125,163,241,236,314,352$},
\quad \mbox{and} \quad  s_{ikl}=-s_{ilk}\ ,\nn \\
&& S_{iklj} =1 \quad \mbox{for
$iklj= 2451,1245,4512,5124,1643,6431,3164,4316,3562,2356,5623,6235$} \ .
\nn
\een
We have written our result in this form, so to be clear as much
as possible the separation between the deformed and the undeformed
part of the Lagrangian.

Following the derivation of \cite{dgo}, one can derive the form of the
deformed Wilson loop. What is relevant is the eigenvalues of
the mass matrix \eq{wfinal}. In the undeformed case, the mass matrix has an
eigenvalue which is 5-fold degenerated and a zero non-degenerate
eigenvalue. The supersymmetric
Wilson loop \eq{wilsono}, \eq{basicconstr} is derived
from the (infinitely) massive quark probe. In the $\b$-deformed case,
the eigenvalues are generally deformed and degeneracy is lifted.
However it is clear that the large $N$
Wilson loop  will be the same as in the undeformed
case because
there isn't any multiplicative factor depending on $N$ in the mass matrix
\eq{wfinal}, therefore
the classical
Lagrangian is the same as the undeformed one in the large $N$ limit \eq{largeN}.

For finite $N$, one will need to keep track of all the dependence of $\b$
in the Lagrangian \eq{wfinal}.  Due to the
large amount of computational work, we were not able to
work out the explicit expressions of the eigenvalues.
However for the cases we have checked (for example by setting some of
the $\phi_k$ and $\th_k$ zero),
it appears that there is always an eigenvalue which is equal to the
undeformed one. It is the phase factor which is associated with
this quark which gives rises to the
Wilson loop \eq{wilsono}, \eq{basicconstr}.

We remark that one may also utilize the star product \eq{star} and
use a star product path ordering to define the Wilson loop operator.
Unlike the Wilson loop in the ordinary noncommutative geometry which
is highly non-local \cite{nc-wil}, the closed Wilson loop
operator is immediately local and there is no need to employ an
open Wilson line. When one expands the exponent, one will get higher
and higher powers of the scalar fields and each of them is accompanied with
a phase factor which depends on the charge configuration of the
scalars. Since these phase factors becomes higher and higher power in $\b$,
in general one cannot drop the $\beta$-dependence even in the large $N$
limit. This operator is not
what one obtains from the probe analysis presented above. It is an
interesting question whether this noncommutative Wilson loop also
admits a nice holographic interpretation, and how.

\section{The deformed metric in the Cartesian coordinate system}

For convenience we collect and present the metric in the coordinate
system
\eqref{yy}
expressed in $Y^i$ coordinates. Defining
\ben
\nonumber
&&A_1=1+\hgamma^2 Y^{-4}(Y^{2^2}+Y^{5^2})(Y^{3^2}+Y^{6^2}),\\
\nonumber
&&A_2=1+\hgamma^2 Y^{-4}(Y^{1^2}+Y^{4^2})(Y^{3^2}+Y^{6^2}), \\
&&A_3=1+\hgamma^2 Y^{-4}(Y^{1^2}+Y^{4^2})(Y^{2^2}+Y^{5^2}),
\een
the metric elements are:
\ben\label{gY11}\nonumber
&&G_{11}=Y^{-2}\frac{(Y^{1^2}+G Y^{4^2} A_1)}{Y^{1^2}+ Y^{4^2}},\quad
G_{44}=Y^{-2}\frac{(Y^{4^2}+G Y^{1^2} A_1)}{Y^{1^2}+ Y^{4^2}},\\
\nonumber
&&G_{22}=Y^{-2}\frac{(Y^{2^2}+G Y^{5^2} A_2)}{Y^{2^2}+ Y^{5^2}},\quad
G_{55}=Y^{-2}\frac{(Y^{5^2}+G Y^{2^2} A_2)}{Y^{2^2}+ Y^{5^2}},\\
&&G_{33}=Y^{-2}\frac{(Y^{3^2}+G Y^{6^2} A_3)}{Y^{3^2}+ Y^{6^2}},\quad
G_{44}=Y^{-2}\frac{(Y^{6^2}+G Y^{3^2} A_1)}{Y^{3^2}+ Y^{6^2}},
\een
\ben\label{gY22}
&& \nonumber G_{12}=\,\,\,\,2 Y^{-6}\hgamma^2 G (Y^{3^2}+
Y^{6^2})Y^{4}Y^5,\,\quad  G_{13}=\,\,\,\,2 Y^{-6}\hgamma^2 G (Y^{2^2}+
Y^{5^2})Y^4Y^6, \, \\
\nonumber
&&G_{15}=-2 Y^{-6}\hgamma^2 G (Y^{3^2}+ Y^{6^2})Y^2 Y^4,\quad G_{16}=-2
Y^{-6}\hgamma^2 G (Y^{2^2}+ Y^{5^2})Y^4Y^3, \, \\
\nonumber
&&G_{23}=\,\,\,\,2 Y^{-6}\hgamma^2 G (Y^{1^2}+ Y^{4^2})Y^5Y^6,\,\quad
G_{24}=-2 Y^{-6}\hgamma^2 G (Y^{3^2}+ Y^{6^2})Y^1Y^5, \, \\   \nonumber
&&G_{26}=-2 Y^{-6}\hgamma^2 G (Y^{1^2}+ Y^{4^2})Y^3Y^5,\quad G_{34}=-2
Y^{-6}\hgamma^2 G (Y^{2^2}+ Y^{5^2})Y^1Y^6,
\\
\nonumber
&& G_{35}=- 2 Y^{-6}\hgamma^2 G (Y^{1^2}+ Y^{4^2})Y^2Y^6,\,\quad
G_{45}=\,\,\,\, 2 Y^{-6}\hgamma^2 G (Y^{3^2}+ Y^{6^2})Y^1Y^2, \, \\
&&G_{46}=\,\,\,\,2 Y^{-6}\hgamma^2 G (Y^{2^2}+ Y^{5^2})Y^1Y^3,\,\,\quad
G_{56}=\,\,\,\,2 Y^{-6}\hgamma^2 G (Y^{1^2}+ Y^{4^2})Y^2Y^3,
\een
\bea \label{gY33}
G_{14}=2 Y^{-2}\frac{Y^1 Y^4(1-G A_1)}{Y^{1^2}+ Y^{4^2}},\nn\\
G_{25}=2 Y^{-2}\frac{Y^2 Y^5(1-G A_2)}{Y^{2^2}+ Y^{5^2}},\\
G_{36}=2 Y^{-2}\frac{Y^3 Y^6(1-G A_3)}{Y^{3^2}+ Y^{6^2}}. \nn
\eea
Substituting from \eqref{yy} the coordinates we express the
metric in angles, the diagonal terms are
\ben\label{g11}\nonumber
&&G_{11}=\frac{1}{Y^2}(\cos^2 \phi_1 + G\sin^2\phi_1 M_1),\quad
G_{44}=\frac{1}{Y^2}(\sin^2 \phi_1 + G\cos^2\phi_1 M_1),\\
\nonumber
&&G_{22}=\frac{1}{Y^2}(\cos^2 \phi_2 + G \sin^2\phi_2 M_2),\quad
G_{55}=\frac{1}{Y^2}(\sin^2 \phi_2 + G \cos^2\phi_2 M_2),\\
&&G_{33}=\frac{1}{Y^2}(\cos^2 \phi_3 + G \sin^2\phi_3 M_3),\quad
G_{66}=\frac{1}{Y^2}(\sin^2 \phi_3 + G \cos^2\phi_3 M_3) \ .
\een
The non-diagonal elements are
\ben\label{g22}
&& \nonumber G_{12}=\,\,\,\frac{1}{Y^2}\hgamma^2 G \mu_1 \mu_2 \mu_3^2
\sin\phi_1 \sin\phi_2,\,\,\quad  G_{13}=\,\,\,\,\frac{1}{Y^2}\hgamma^2
G \mu_1 \mu_2^2 \mu_3  \sin\phi_1 \sin\phi_3, \, \\
\nonumber
&&G_{15}=-\frac{1}{Y^2}\hgamma^2 G \mu_1 \mu_2 \mu_3^2 \sin\phi_1
\cos\phi_2,\quad G_{16}=-\frac{1}{Y^2}\hgamma^2 G \mu_1 \mu_2^2 \mu_3
\sin\phi_1 \cos\phi_3, \, \\
\nonumber
&&G_{23}=\,\,\,\frac{1}{Y^2}\hgamma^2 G \mu_1^2 \mu_2 \mu_3 \sin\phi_2
\sin\phi_3,\,\,\quad
G_{24}=-\frac{1}{Y^2}\hgamma^2 G \mu_1 \mu_2 \mu_3^2 \cos\phi_1
\sin\phi_2, \, \\   \nonumber
&&G_{26}=-\frac{1}{Y^2}\hgamma^2 G \mu_1^2 \mu_2 \mu_3 \sin\phi_2
\cos\phi_3,\quad G_{34}=-\frac{1}{Y^2}\hgamma^2 G \mu_1 \mu_2^2 \mu_3
\cos\phi_1 \sin\phi_3,
\\
&& \nonumber G_{35}=-\frac{1}{Y^2}\hgamma^2 G \mu_1^2 \mu_2 \mu_3
\cos\phi_2 \sin\phi_3,\,\quad G_{45}=\,\,\,\frac{1}{Y^2}\hgamma^2 G
\mu_1 \mu_2 \mu_3^2 \cos\phi_1 \cos\phi_2, \, \\
&&G_{46}=\,\,\,\frac{1}{Y^2}\hgamma^2 G \mu_1 \mu_2^2 \mu_3 \cos\phi_1
\cos\phi_3,\,\,\quad
G_{56}=\,\,\,\frac{1}{Y^2}\hgamma^2 G \mu_1^2 \mu_2 \mu_3 \cos\phi_2
\cos\phi_3  ,
\een
and
\ben \label{g33}
G_{14}=\frac{1}{2Y^2}\sin2 \phi_1(1-G M_1),\,
G_{25}=\frac{1}{2Y^2}\sin 2 \phi_2(1-G M_2),\,
G_{36}=\frac{1}{2Y^2}\sin 2 \phi_3(1-G M_3).\;\;\;\; 
\een

\section{Derivation of the Hamilton-Jacobi Equation}
In this appendix we shortly derive the Hamilton-Jacobi (HJ) equation
(\ref{HJ0}).
Consider the action for the string
\be
S= \int d^2 \s \bigl( \sqrt{\det g} -i   B_{IJ} \del_1 X^I \del_2 X^J \bigr)
\ee
where $g_{\a\b} := G_{IJ} \del_\a X^I \del_\b X^J$, $\a,\b
=1,2$. The conjugate momentum is
\be
P_I = \frac{\d S}{\d (\del_2 X^I)} = \frac{1}{\sqrt{g}} 
G_{IJ} (g_{11} \del_2 X^J -g_{12} \del_1 X^J) +i B_{IJ} \del_1 X^J
:= \cP_I +i B_{IJ} \del_1 X^J, 
\ee
where we have introduced $\cP_I$ as defined above. This turns out to
be a convenient variable for expressing the HJ equation.
The Hamiltonian is
\be
H= P_I \del_2 X^I -L = \cP_I \del_2 X^I - \sqrt{g}.
\ee
Eliminate $\del_2 X^I$  in terms of $\cP_I$ and note that $\cP_I
\del_1 X^I =0$, we obtain
\be
H = \frac{\sqrt{g}}{g_{11}} (G^{IJ} \cP_I \cP_J - g_{11}). 
\ee
And we obtain the HJ equation $H=0$,
\be \label{HJ}
G^{IJ} \cP_I \cP_J = G_{IJ} \del_1 X^I \del_1 X^J.
\ee
This is the form of HJ equation we used in the main text of the paper.

\section{Cancellation of UV divergences up to order $(g^2 N)^2$ }
  
We first demonstrate that
that the scalar propagator and the
gauge boson propagator in the Feynman gauge remains equal up to first
order in $g^2 N$. The simplest way to show this is to use superspace
Feynman graphs. In terms of superfields, the Lagrangian for the
$\b$-deformed SYM theory is
\bea
L &=& \int d^2 \th d^2 \thb \; \Tr(e^{-gV} \Phib_i e^{gV} \Phi_i) +
\frac{1}{2 g^2} \int d^2 \th \; \Tr W^\a W_\a +\mbox{c.c.}\\
&+& 
i h \int d^2 \th \; \Tr(e^{i \pi \b} \Phi_1 \Phi_2 \Phi_3 - e^{-i \pi
  \b} \Phi_1\Phi_3 \Phi_2)
+ i h^* \int d^2 \thb \; \Tr(e^{i \pi \b} \Phib_1 \Phib_2 \Phib_3 - e^{-i \pi
  \b} \Phib_1\Phib_3 \Phib_2). \nn
\eea
Using
$ f_{abc} := -i \Tr(T_a[T_b,T_c]) , \quad d_{abc} := \Tr(T_a \{T_b,T_c\})$,
the superpotential can be written as
\be
-h (f_{abc} \cos\pi \b + d_{abc} \sin \pi \b) \int d^2 \th \;
\Phi^a_1 \Phi^b_2 \Phi^c_3 + \mbox{c.c.} \; .
\ee
The relation between $h$ and $g$ is obtained from the requirement of
superconformal invariance, which gives up to two-loop order
\cite{dzf,z1},
\be \label{scf}
 |h|^2 \big(C_2 \cos^2 \pi \b +  D_2 \sin^2 \pi \b \big) = N g^2.
\ee
Here $f_{abc} f_{a'bc} = \d_{aa'} C_2$, $d_{abc} d_{a'bc}  =\d_{aa'}
D_2$ and $\Tr(T_a T_{a'}) =\d_{aa'}/2$. 
\if 
It is $C_2 =D_2=N$ for $U(N)$
and  for $SU(N)$: $C_2 =N$, $D_2 = (N^2-4)/N$. This gives
\bea
\mbox{$U(N)$:} &&\quad h=g , \\
\mbox{$SU(N)$:} && \quad
|h|^2\bigg( \cos^2 \pi \b + \big(1-\frac{4}{N^2}\big)\sin^2 \pi \b \bigg) =g^2.
\eea
\fi
Now the 1-loop correction to the scalar propagator is contained in the
diagrams in figure 1.
\begin{figure}
\label{fig1}
\begin{center}
{\scalebox{0.7}{ \includegraphics{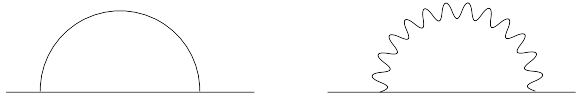}} }
\caption{1-loop contribution to scalar propagator}
\end{center}
\end{figure}
It is obvious that the graph (b) is independent of $\b$. For
the graph (a), it has a interaction vertex proportional to
$ |h|^2 (f_{abc} f_{a'bc} \cos^2 \pi \b + d_{abc} d_{a'bc} 
\sin^2 \pi \b). $ 
Using the superconformal invariance condition \eq{scf}, 
this is equal to $g^2 N \d_{aa'}$ and is independent of $\b$. Thus the one loop
contribution to the scalar propagator is independent of $\b$. It is
obvious that the  one loop
contribution to the gauge boson propagator is also independent of $\b$. Using
the result of \cite{ESZ}, we conclude that the scalar propagator and the
gauge boson propagator remains equal up to first order in $g^2 N$.


Using this result, it is easy to see that the Wilson
loop operator \eq{wilsono} is free from UV divergence up to order
$(g^2 N)^2$ if the constraint \eq{basicconstr} is satisfied. 
The proof is a slight adaption of the computation of \cite{zarembo}.
At  leading and next-to-leading orders, we have the
Feynman diagrams given in figure 2.
\begin{figure}
\label{fig2}
\begin{center}
{\scalebox{0.9}{ \includegraphics{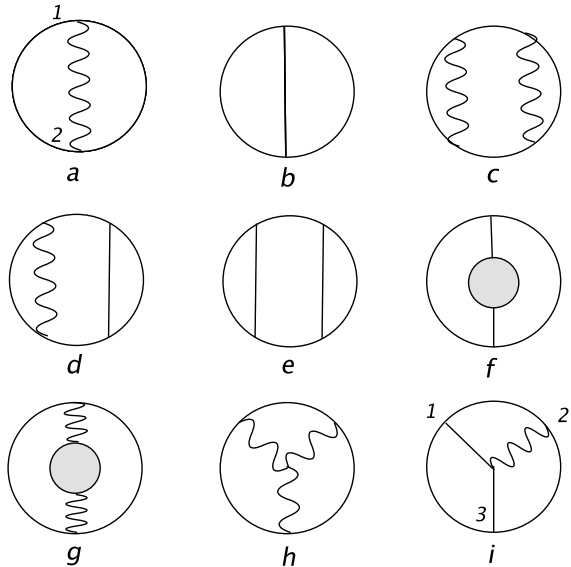}} }
\caption{Feynman diagrams of leading and next-to-leading orders}
\end{center}
\end{figure}
\if
The diagrams (a) and (b) has a linear divergence when point 1 is
getting close to 2. The divergence is  cancelled out when the
constraint holds \cite{dgo}. Similarly the linear divergence from the diagrams
(c,d,e) cancels. Due to the equality of the 1-loop corrected
scalar and gauge boson
propagators, the linear divergence from diagrams (f,g) also
cancels. 
\fi
The linear divergences in diagrams (a-g) got cancelled out immediately
due  to the equality of the 1-loop corrected
scalar and gauge boson propagators.
As for the diagrams (h) and (i),  we have 
\bea
(h)+(i) &=&
\if
\frac{2}{g^2N} \int d^4 x \oint ds_1 ds_2 ds_3
\th_c(s_1,s_2,s_3) \\
&&\times \bigg(
\langle \frac{1}{3} \Tr A_\mu(x(s_1))A_\nu(x(s_2)) A_\l(x(s_3))
\Tr\del_\rho A_\s(x)[A_\rho(x),A_\s(x)]\rangle
\dot{x}^\m(s_1)\dot{x}^\nu(s_2)\dot{x}^\l(s_3) \nn\\ 
&& 
-\langle \Tr \Phi_i(x(s_1))\Phi_j(x(s_2)) A_\l(x(s_3))
\Tr \del_\rho \Phi_k(x)[A_\rho(x),\Phi_k(x)]\rangle
\dot{y}^i(s_1)\dot{y}^j(s_2)\dot{x}^\l(s_3)
\bigg)\nn\\
&=& 
\fi
2(g^2N)^2 \int d^4 x \oint ds_1 ds_2 ds_3
\th_c(s_1,s_2,s_3) \cdot \nn\\
&&\qquad \qquad\qquad
\cdot (D_{xx_1}\del_\l D_{xx_2} -\del_\l D_{xx_1} D_{xx_2})
D_{xx_3}\dot{x}_3^\l \cdot
(\dot{x}_1^\mu\dot{x}_2^\nu \d_{\m\n} -\dot{y}_1^i\dot{y}_2^j
\d_{ij}). \; \;\;\label{hi2}
\eea
The contribution to \eq{hi2} from the region $s_1 \sim s_2
\sim s_3$ is linear divergent  for a generic loop,
\be
(h)+(i) \sim \oint ds_1 \frac{1}{\e}  (\dot{x}_1^2 - \dot{y}_1^2 + \e) .
\ee
However when the constraint \eq{basicconstr} is satisfied, the
contribution is finite. Thus we conclude that the Wilson loop operator
\eq{wilsono} has a expectation value that is free from UV divergence
up to order $(g^2 N)^2$ when the constraint is satisfied.

We also remark that, due to the equality of the propagators, 
the Wilson loop operator with the constraint \cite{zarembo}
\be
\dot{y}^i = M^i_\m \dot{x}^\m, \quad M^i_\m M^i_\n = \d_{\m\n}
\ee
has expectation value 1 up to order  $(g^2 N)^2$. We conjecture that this
Wilson loop has an exact expectation value 1 just as in the $\cN=4$ theory.

\if
These kind of Wilson loop operators were first
introduced by Zarembo \cite{zarembo} for the $\cN=4$ theory.
They are supersymmetric and
the amount of preserved supersymmetry depends on the dimensionality of
the loop. In our case of $\b$-deformed SYM, they are
not supersymmetric anymore. However one can perform
the same  analysis as in \cite{zarembo}. Since the 1-loop corrected
propagators take the same form as in the undeformed case, and since the
relevant vertices are undeformed, one concludes  immediately that all
corrections of order $g^2 N$ and $(g^2 N)^2$ cancel. We conjecture that this
Wilson loop has expectation value 1 just as in the undeformed case.
\fi

\end{document}